# Superconductivity at 10.4 K in a novel quasi-one-dimensional ternary molybdenum pnictide K$_2$Mo$_3$As$_3$

Qing-Ge Mu[1,2], Bin-Bin Ruan[1,2], Kang Zhao[1,2], Bo-Jin Pan[1,2], Tong Liu[1,2], Lei Shan[1,2,3], Gen-Fu Chen[1,2,3], and Zhi-An Ren[1,2,3,*]

[1] Institute of Physics and Beijing National Laboratory for Condensed Matter Physics, Chinese Academy of Sciences, Beijing 100190, China

[2] School of Physical Sciences, University of Chinese Academy of Sciences, Beijing 100049, China

[3] Collaborative Innovation Center of Quantum Matter, Beijing 100190, China

[*] Email: renzhian@iphy.ac.cn

**Abstract**

Here we report the discovery of the first ternary molybdenum pnictide based superconductor K$_2$Mo$_3$As$_3$. Polycrystalline samples were synthesized by the conventional solid state reaction method. X-ray diffraction analysis reveals a quasi-one-dimensional hexagonal crystal structure with (Mo$_3$As$_3$)$^{2-}$ linear chains separated by K$^+$ ions, similar as previously reported K$_2$Cr$_3$As$_3$, with the space group of $P$-6$m$2 (No. 187) and the refined lattice parameters $a$ = 10.145(5) Å and $c$ = 4.453(8) Å. Electrical resistivity, magnetic susceptibility, and heat capacity measurements exhibit bulk superconductivity with the onset $T_c$ at 10.4 K in K$_2$Mo$_3$As$_3$ which is higher than the isostructural Cr-based superconductors. Being the same group VIB transition elements and with similar structural motifs, these Cr and Mo based superconductors may share some common underlying origins for the occurrence of superconductivity and need more investigations to uncover the electron pairing within a quasi-one-dimensional chain structure.

1. **Introduction**

In regardless of the unclear underlying mechanism for the unconventional high-$T_c$ superconductivity in cuprates and iron pnictides/chalcogenides, the occurrence of superconductivity is strongly related to the existence of some certain structural motifs as $CuO_2$ planes or $Fe_2As_2$/$Fe_2Se_2$ layers in varied crystal structures, which actually has been the guidance for searching new superconductors in these materials [1-6]. Back to the early 1970's, there were a large number of molybdenum chalcogenide superconductors $M_xMo_6X_8$ (X = S, Se, or Te) discovered, also known as the Chevrel phases, in which M can be diverse metallic elements with x varying from 0 to 4 [7-17]. The basic structural motif for these compounds is the $Mo_6X_8$ cluster, in which six Mo atoms forms an octahedron and eight chalcogenide atoms are connected to the triangular facets from outside, and other types of metallic atoms can be inserted between the clusters to constitute versatile materials. When the $Mo_6X_8$ clusters are linearly connected together and condensed into an infinite monodimensional chain structure, a new series of compounds $M_2Mo_6X_6$ were discovered, and some of them are superconducting with quasi-one-dimensional (Q1D) features [18-22]. These Chevrel phase superconductors were regarded as remarkable high temperature superconductors due to their high $T_c$ and high upper critical field and attracted intensive research interests in those years (for $PbMo_6S_8$, $T_c$ ~ 15 K, $H_{c2}$ ~ 60 T) [9, 23]. Some of them exhibit more exotic physical phenomena such as the reentrant superconducting phase, the coexistence of superconductivity with long-range ferromagnetic order in $HoMo_6S_8$, and the unique magnetic field induced superconductivity in $Eu_{0.75}Sn_{0.25}Mo_6S_{7.2}Se_{0.8}$ [10, 16]. However, the study on these superconductors was almost broken off by the later discovery of high-$T_c$ cuprates and iron-based superconductors in spite of many fundamental problems are still under debate [17].

Recently, a novel family of chromium arsenide based Q1D superconductors $A_2Cr_3As_3$ (A = alkali metal) were reported, which shares the same $(Cr_3As_3)^{2-}$ chain-like structural motif just as the $(Mo_6X_6)^{2-}$ [24-27]. These Cr-233 superconductors crystallize in a noncentrosymmetric hexagonal structure with Q1D $(Cr_3As_3)^{2-}$ linear chains separated by alkali metal cations. Strong electron correlations and magnetic fluctuations, highly anisotropic upper critical field exceeding the Pauli-pair breaking limit, line nodes in the superconducting gap, and Tomonaga-Luttinger liquid behavior were observed by different experimental techniques and showed the possibility of unconventional superconductivity [24-41]. Spin-triplet electron pairing was proposed in these Cr-233 superconductors from the beginning but a consensus is still far to be concluded [24-34, 38, 39]. And lately, superconductivity was also found in the Q1D-type $ACr_3As_3$ crystals which actually have identical crystal lattice to the Mo-based $M_2Mo_6X_6$ [18, 42-45]. These 133-type Cr-based superconductors exhibit similar high upper critical field and strong electron correlations within a centrosymmetric lattice, but negative chemical pressure effect on $T_c$ from the replacement of alkali metal cations which is opposite to that of the Cr-233 superconductors, and more investigations are still ongoing to further understand their physical characteristics [24-27, 44, 45].

As the elements Cr and Mo belong to the same group VIB transition metals with similar electronic configuration, a question arising is whether these two families of superconductors share some common underlying origins for the occurrence of superconductivity.

In this manuscript, we report the successful synthesis of a Q1D compound $K_2Mo_3As_3$ that has the same crystal structure as the Cr-233 superconductors. This Mo-233 type $K_2Mo_3As_3$ shows bulk superconductivity at a higher $T_c$ of 10.4 K, which may shed new lights on the understanding of the superconducting orders in these group VIB transition metal compounds.

2. **Experimental details**

Polycrystalline $K_2Mo_3As_3$ samples were synthesized by the conventional solid state reaction method in two steps, using elemental K pieces (99%), Mo powder (99.95%) and As powder (99.999%) as the starting materials. At first, the mixtures of K, Mo and As elements with the atomic ratio of 2.5:3:3 were placed in an alumina crucible, covered by an alumina cap and then sealed in an evacuated quartz tube (~ $10^{-4}$ Pa). The sealed quartz tube was slowly heated to 523 K and kept for 20 h in a muffle furnace, and then cooled down to room temperature. After the initial reaction, the mixture of these intermediate products could be thoroughly ground into fine powders, which were then pressed into small pellets. Secondly, the pellets were placed into an alumina crucible, and sealed into a Ta tube with arc welding in argon atmosphere (~ 0.05 MPa). The Ta tube was sealed in an evacuated quartz tube, and then it was slowly heated to 1123 K and sintered for 50 h in the muffle furnace before cooled down to room temperature by furnace shut-down. The preparation procedures were strictly carried out in a glove box filled with high-purity Ar gas ($O_2$ and $H_2O$ content less than 0.1 ppm) to avoid any possible contamination by oxygen or moisture to the samples. Due to the easy volatilization of alkali metals and arsenic, to obtain single phase samples with no obvious binary impurities, we did many attempts for the optimization of the synthesizing conditions and the ratio of starting materials since no ternary chemical phase of alkali metal molybdenum arsenide has ever been reported.

The obtained $K_2Mo_3As_3$ samples are black in color and extremely reactive in air, hence any exposure to air should be avoided when performing measurements on these samples. Needle-like tiny crystals are occasionally observed on the as-grown sample surface. The crystal structure was characterized at room temperature by powder x-ray diffraction (XRD) using a PAN-analytical x-ray diffractometer with Cu-$K_\alpha$ radiation. The sample morphology was characterized with a Phenom scanning electron microscope (SEM). The electrical resistivity and heat capacity measurements were performed on a Quantum Design physical property measurement system with the standard four-probe method and thermal relaxation method, respectively. The dc magnetization was measured with a Quantum Design magnetic property measurement system.

### 3.    Results and discussion

The crystal structure of $K_2Mo_3As_3$ is determined by the powder XRD analysis, which reveals identical lattice symmetry with that of previously reported Q1D noncentrosymmetric $K_2Cr_3As_3$-type superconductors [24]. In Fig. 1(a) the crystal structure is depicted with the schematic $M_6X_8$ cluster and the $M_3X_3$ chain condensed from the cluster. The $K_2Mo_3As_3$ can be considered as linear hexagonally arranged $(Mo_3As_3)^{2-}$ chains separated by the $K^+$ alkali cations that act as a charge reservoir. The SEM morphology characterizations for the fresh fracture surface of the as-grown sample show all needle-like crystal grains and explicitly manifest the Q1D lattice structure, as displayed in Fig. 1(b). The XRD analysis was performed on powdered samples and the typical diffraction pattern with $2\theta$ from 5º to 90º is shown in Fig. 1(c). All the reflection peaks can be well indexed by the hexagonal lattice structure of the space group *P*-6*m*2 (No. 187) with no impurity phase observed, and the refined lattice parameters are $a$ = 10.145(5) Å and $c$ = 4.453(8) Å. We note that due to the superior ductility for $K_2Mo_3As_3$ and the highly orientation of the needle-like crystal grains, the intensity for the diffraction peaks is not well refined, hence we have not obtained the detailed atomic coordinates from the diffraction data. Comparing with that of the isostructural $K_2Cr_3As_3$, the crystal lattices expand more evidently along *c*-axis (~ 5.3%) than *a*-axis (~ 1.6%) due to the replacement of transition metals, which is quite different with the replacement of alkali metals in the $K_2Cr_3As_3$ series that mainly alters the *a*-axis parameter or the inter-chain distance [24-27]. This suggests that the Q1D chain structure is dominated by the transition metal elements bonded with arsenic, while the inter-chain bonding is much weaker. Here we note that the excess using of elemental K in the starting materials is important to obtain single-phase sample, otherwise lots of binary impurities would be observed in the sintered sample.

As we measured the low temperature resistivity for $K_2Mo_3As_3$, superconducting transitions were observed in all batches of samples. In Fig. 2(a) we show the temperature dependence of electrical resistivity for three typical samples of $K_2Mo_3As_3$ from 1.8 K to 300 K under zero fields. The electrical resistivity $\rho(T)$ usually exhibits metallic characteristic in the normal state while semiconducting behavior sometimes appears at low temperature in some batches of samples. High quality single crystal is necessary to clarify the normal state electrical transport characteristics. All samples exhibit unambiguous superconducting transitions near the similar onset critical temperature of $T_c$ ~ 10.4 K with zero resistance appeared at low temperature, and the typical superconducting transition width is about 0.3 K as shown in the inset for sample A. The $T_c$ in this Q1D type $K_2Mo_3As_3$ compound remarkably exceeds that of all other previously reported Q1D superconductors [18, 20, 21, 24-27, 44-48]. To characterize the upper critical field $\mu_0H_{c2}$, the electrical resistivity from 2 K to 12 K with temperature sweeping was measured for sample A under a constant magnetic field, the fields were varied from 0 T to 16 T with 1 T interval, and the data are shown in the inset of Fig. 2(b). Upon applying magnetic field, $T_c$ shifts to lower temperature sharply at first, and then it becomes slowly and the superconducting transition width shows a broadening effect similar as $Tl_2Mo_6Se_6$ [20]. In the normal state, no obvious magnetoresistance effect appears. Figure 2(b) represents the temperature dependence of $\mu_0H_{c2}$, and the data show a slowly increase of the absolute value for the $dH_{c2}/dT$ when temperature decreases. The data are fitted with the Ginzburg-Landau formula, $H_{c2}(T) = H_{c2}(0)(1 - t^2)/(1 + t^2)$, in which $t = T/T_c$. The zero-temperature upper critical field $\mu_0H_{c2}(0)$ is estimated to be 22.0 T, which is just above the Pauli paramagnetic limited critical field $\mu_0H_p = 1.84T_c = 19.1$ T [49]. Moreover, the behavior of the $dH_{c2}/dT$ does not exhibit obvious Pauli pair breaking effect under the measured fields, which may indicate possible unconventional superconductivity in $K_2Mo_3As_3$.

To demonstrate the bulk superconductivity in $K_2Mo_3As_3$, the temperature dependence of dc susceptibility and heat capacity were measured and shown in Fig. 3. In Fig. 3(a) we show the susceptibility data for the above-mentioned three samples from 2 K to 12 K with zero-field-cooling (ZFC) and field-cooling (FC) modes under a stable magnetic field of 10 Oe. Both ZFC and FC data show clear diamagnetic superconducting transitions below 10.2 K, which is consistent with the electrical resistivity measurements. The diamagnetic shielding volume fraction derived from the ZFC data is close to 100% at 2 K, suggesting the bulk superconductivity in $K_2Mo_3As_3$. The isothermal magnetization measurement at $T = 2$ K for sample A exhibits typical type-II superconductivity as shown in Fig. 3(b). The high-temperature susceptibility $\chi(T)$ follows Curie-Weiss behavior with no magnetic order appearing. The temperature dependence of heat capacity for sample A is shown as the relationship of $C_p/T$ vs $T^2$ in Fig. 3(c) with a clear superconducting transition. The normal state data are linearly fitted with both electron and phonon contributions by $C_p/T = \gamma + \beta T^2$. From the fitted parameters we get the Sommerfeld coefficient $\gamma$ as 13 mJ mol$^{-1}$ K$^{-2}$ which indicate much weakened electron correlations than Cr-233 superconductors [24-27], and the Debye temperature $\theta_D$ as 234 K calculated from $\theta_D = [(12/5)NR\pi^4/\beta]^{1/3}$. The temperature dependence of electron contributions for heat capacity is normalized as $(C - C_{ph})/(\gamma T)$ vs. $T$ and shown in Fig. 3(d) (where $C_{ph}$ is the phonon contribution of $\beta T^3$), which exhibits the characteristic heat capacity jump ($\Delta C$) for the superconducting transition. The dimensionless heat capacity jump $\Delta C/\gamma T$ at $T_c$ is about 1.56, which confirms the bulk superconductivity in $K_2Mo_3As_3$. We note that the value of $\Delta C/\gamma T$ is smaller for sample C, which might be due to possible variation of sample stoichiometry that affects the occurrence of superconductivity severely.

In conclusion, we successfully synthesized a MoAs-based ternary compound $K_2Mo_3As_3$ which has a typical Q1D crystal structure by the conventional solid state reaction method. Electrical resistivity, dc magnetization and heat capacity measurements revealed the existence of bulk superconductivity in $K_2Mo_3As_3$ with an onset $T_c$ of 10.4 K. This discovery provides the first MoAs-based superconductor and may help to uncover more insights into the deep physics of Cr and Mo based Chevrel phases. Note added: The isostructural $Rb_2Mo_3As_3$ and $Cs_2Mo_3As_3$ were also synthesized with superconductivity observed at 10.6 K and 11.5 K respectively.

**Acknowledgements**

The authors are grateful for the financial supports from the National Natural Science Foundation of China (No. 11474339 and 11774402), the National Basic Research Program of China (973 Program, No. 2016YFA0300301) and the Youth Innovation Promotion Association of the Chinese Academy of Sciences.

**Figure captions:**

Fig. 1. (a) The schematic crystal structure for the $M_6X_8$ cluster, the $M_3X_3$ chain, and the $K_2Mo_3As_3$ lattice. (b) The SEM morphology characterization for the fresh fracture surface of the $K_2Mo_3As_3$ polycrystalline sample. (c) The powder XRD patterns and structural refinement for $K_2Mo_3As_3$.

Fig. 2. (a) The temperature dependence of electrical resistivity for three typical polycrystalline $K_2Mo_3As_3$ samples; and the inset depicts the superconducting transition for sample A. (b) The temperature dependence of the derived upper critical field $\mu_0H_{c2}$ for sample A with GL-fit, and the inset depicts the superconducting transitions for sample A under varied magnetic fields up to 16 T.

Fig. 3. (a) The temperature dependence of dc magnetic susceptibility for the three samples of $K_2Mo_3As_3$. (b) The isothermal magnetization curve at 2 K for sample A. (c) The low-temperature heat capacity depicted as $C_p/T$ vs $T^2$ with linear fit for the normal state data. (d) The normalized temperature dependence of electron contributions for heat capacity $(C - C_{ph})/(\gamma T)$ that shows the superconducting jump.

Figure 1

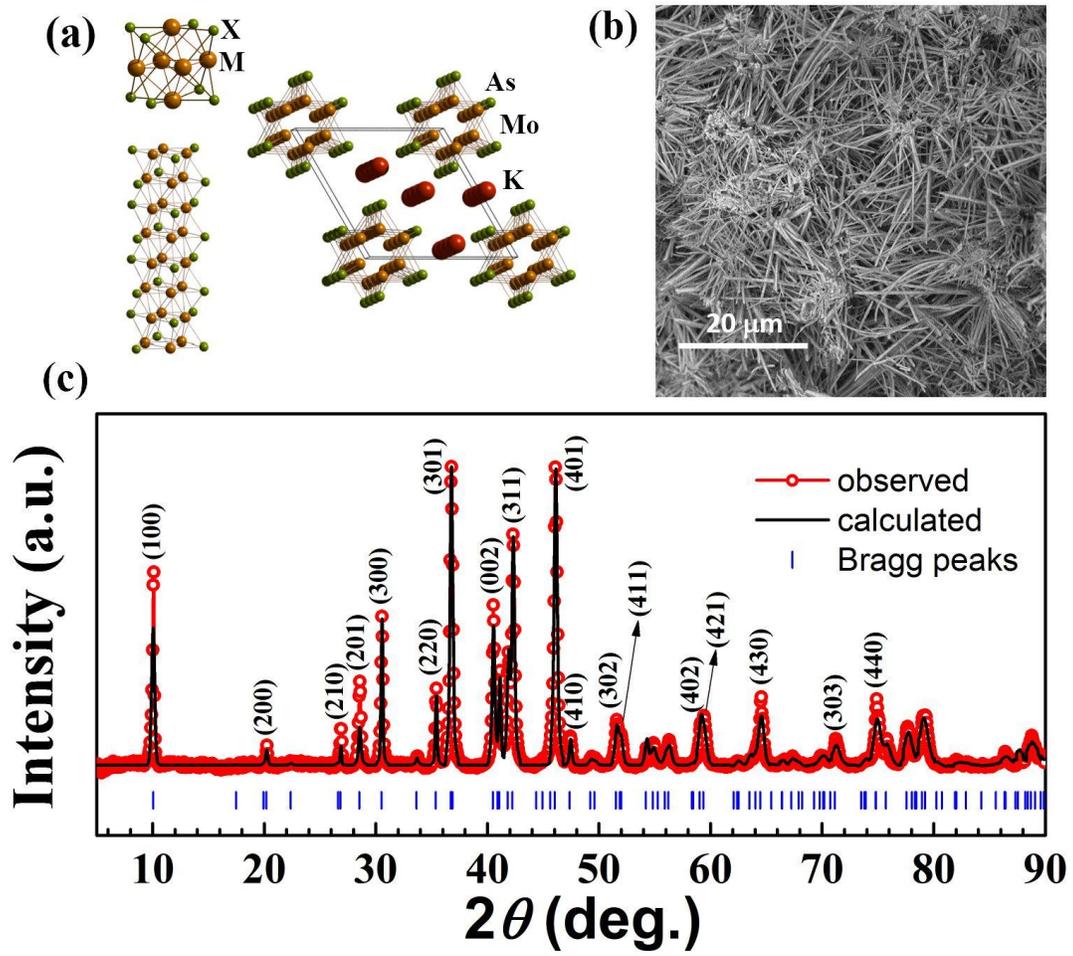

**Figure 2**

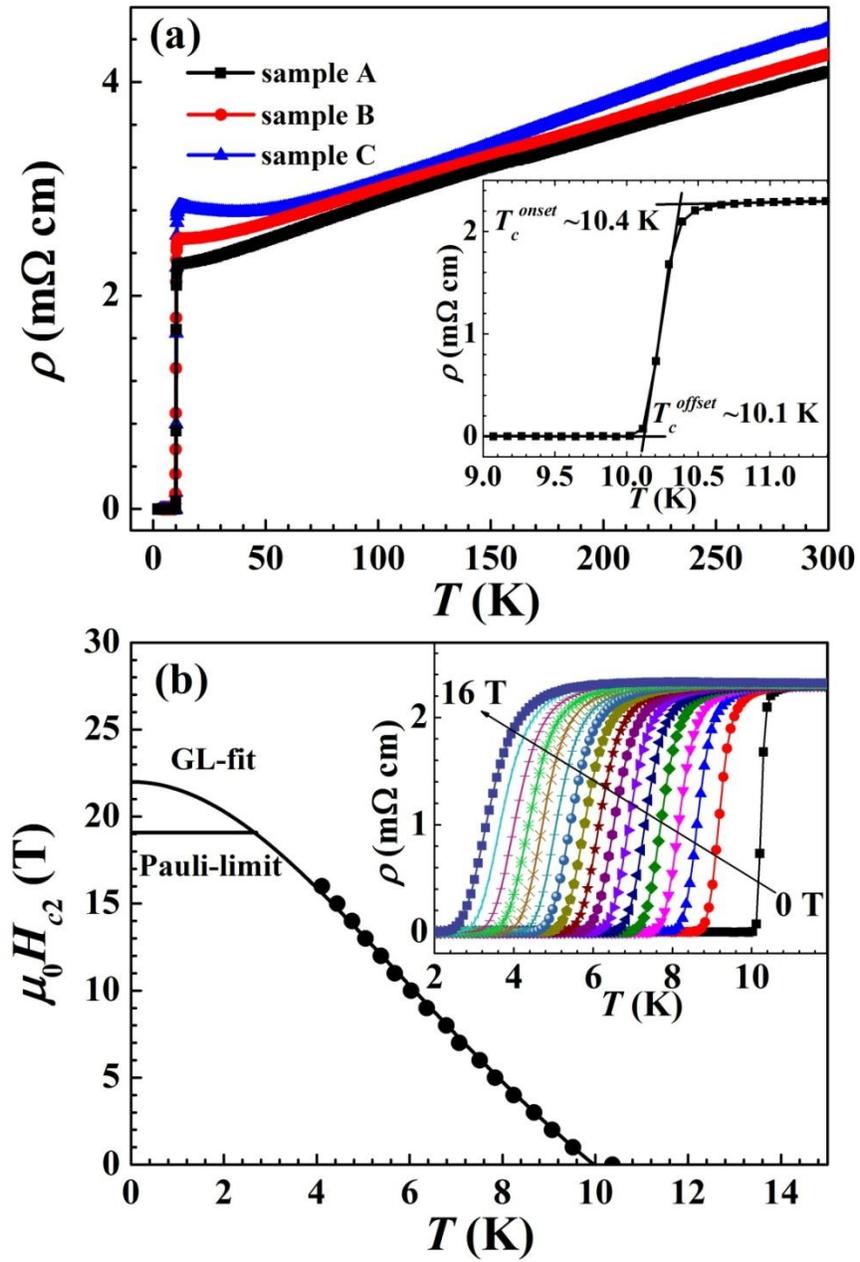

**Figure 3**

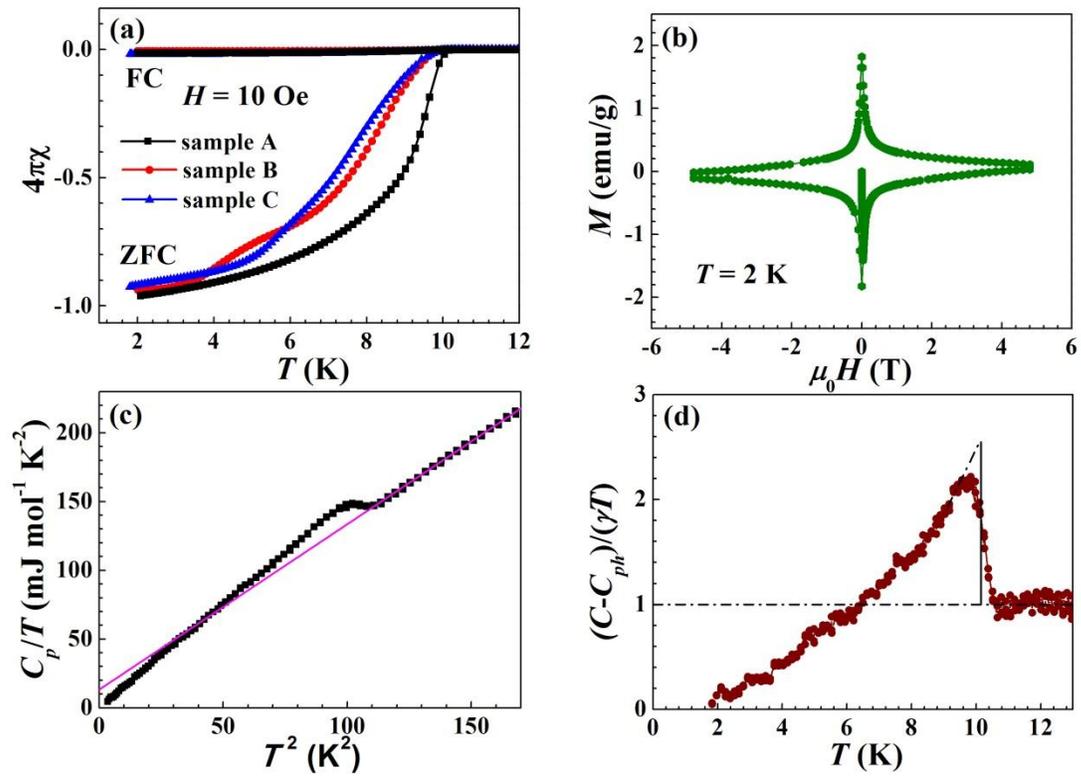